\documentclass[apj,twocolappendix]{emulateapj}

\usepackage{graphicx}
\usepackage{amssymb}
\usepackage{multirow}
\usepackage{float}
\usepackage{comment}

\newcommand{\be}{\begin{equation}}
\newcommand{\ee}{\end{equation}}

\newcommand{\psr}{\hbox{\object{PSR\,B1055$-$52}}}

\begin{document}
 
\title{A {\sl Chandra} search for the pulsar wind nebula around PSR\,B1055$-$52}

\author{B. Posselt}
\affil{Department of Astronomy \& Astrophysics, Pennsylvania State University, 525 Davey Lab,University Park, PA 16802, USA }
\email{posselt@psu.edu}
\author{G. Spence}
\affil{Department of Astronomy \& Astrophysics, Pennsylvania State University, 525 Davey Lab,University Park, PA 16802, USA}
\author{G. G. Pavlov}
\affil{Department of Astronomy \& Astrophysics, Pennsylvania State University, 525 Davey Lab,University Park, PA 16802, USA}

\begin{abstract}
The nearby, middle-aged PSR\,B1055$-$52 has many properties in common with the Geminga pulsar. Motivated by the Geminga's enigmatic and prominent pulsar wind nebula (PWN), we searched for extended emission around PSR\,B1055$-$52 with {\sl{Chandra}} ACIS.
For an energy range $0.3-1$\,keV, we found a $4\sigma$ flux enhancement in a $4\farcs{9}-20\arcsec$ annulus around the pulsar. 
There is a slight asymmetry in the emission close, $1\farcs{5}-4\arcsec$, to the pulsar. 
The excess emission has a luminosity of about $10^{29}$\,erg\,s$^{-1}$ in an energy range $0.3-8$\,keV for a distance of 350\,pc.
Overall, the faint extended emission around $\psr$ is consistent with a PWN of an aligned rotator moving away from us along the line of sight with supersonic velocity, but a contribution from a dust scattering halo cannot be excluded.Comparing the properties of other nearby, middle-aged pulsars, we suggest that the geometry - the orientations of rotation axis, magnetic field axis, and the sight-line -- is the deciding factor for a pulsar to show a prominent PWN.
We also report on an $\gtrsim 30$\,\% flux decrease of PSR\,B1055$-$52 between the 2000 XMM-$Newton$ and our 2012 {\sl{Chandra}} observation.
We tentatively attribute this flux decrease to a cross-calibration problem, but further investigations of the pulsar are required to exclude actual intrinsic flux changes.
\end{abstract}


\section{Introduction}
When pulsars spin down, most of their rotational energy loss is carried away by winds of relativistic particles. When the winds shock in the ambient medium, these shocks emit synchrotron radiation which becomes observable as a pulsar wind nebula ({\rm PWN}, for reviews see, e.g., \citealt{2011Slane,Kargaltsev2008, Gaensler2006}).  
The PWN luminosity $L_{{\rm PWN}}$ correlates with the pulsar spin-down power $\dot{E}$ as $L_{{\rm PWN}}=\dot{E} \times \eta_{\rm PWN}$. 
The efficiencies $\eta_{\rm PWN}$ at X-ray energies are usually between  $\eta_{\rm X} \sim 10^{-5}$ and $\eta_{\rm X} \sim 10^{-3}$ \citep{Kargaltsev2012}.
The wide range of observed efficiencies depend on ages, magnetic fields, environments, and evolutionary states of the neutron stars. 
Older pulsars have smaller $\dot{E}$, their {\rm PWN}e are dimmer and more difficult to observe. Thus, most of the currently known $\sim 60$ {\rm PWN}e are powered by young ($<100$\,kyrs) pulsars, while little is known about PWNe around older pulsars. 
Only very few {\rm PWN}e around middle-aged ($0.1-1$\,Myr) pulsars with $\dot{E}<10^{35}$\,erg\,s$^{-1}$ have been detected.\\

The best known nearby representatives of such middle-aged pulsars are PSR\,B1055$-$52, Geminga and PSR B0656+14. 
They have similar spin-down energies ($\dot{E} \sim 10^{34}$\,erg\,s$^{-1}$), similar ages (a few 100\,kyrs), and similar magnetic fields ($\sim 10^{12}$\,G). All three are $\gamma$-ray pulsars. Their X-ray spectra consist of non-thermal and strong thermal components. Because of the similarities, they were dubbed  the {\textsl{Three Musketeers}} by \citet{Becker1997}.
The X-ray spectra are best described by a three-component model -- a cold and a hot blackbody plus a power law component (e.g., \citealt{Becker2002,Pavlov2002}). \citet{deluca2005} studied the individual components with phase-resolved, low-resolution X-ray spectroscopy.
They found pronounced differences for these three sources indicating different orientations, magnetic field configurations and surface temperature nonuniformities.
The current (June 2015) statistic of nearby ($<2$\,kpc), middle-aged pulsars (characteristic ages between $0.1 -1$\,Myr and $\dot{E}$ between $5\times 10^{33}- 5\times 10^{34}$\,erg\,s$^{-1}$) lists 17 objects (see Table~\ref{pulsars}), many recently discovered with $Fermi$.\\

Putative {\rm PWN}e around the {\textsl{Three Musketeers}}, found with {\sl ASCA} (e.g., \citealt{Shibata1997}), were later resolved into several point sources, thanks to the better spatial resolution of {\sl ROSAT} and {\sl BeppoSAX} \citep{Becker1999}.
Further improvement in spatial resolution and sensitivity resulted in the {\sl XMM-Newton} detection of two long, lateral {\rm PWN}-tails behind Geminga \citep{Caraveo2003}. The superb spatial resolution of {\sl Chandra} revealed the axial tail and the structure of the lateral tails in the immediate vicinity of the Geminga pulsar \citep{Pavlov2010,Pavlov2006}.
Of the 12 middle-aged pulsars (except for \psr) with sensitive, public X-ray observations, 6 have prominent X-ray PWNe, while the other 6 have no detected or very faint PWNe; see Table~\ref{pulsars}.
PSR\,J1741$-$2054, for example, is surrounded by a large H$_{\alpha}$ bow shock nebula and an impressive X-ray PWN \citep{Brownsberger2014, Romani2010}.
PSR\,J0357+3205 has a very long tail, though it was ruled out that this long tail is associated with a classical bow-shock, ram-pressure-dominated PWN (e.g., \citealt{Marelli2013}). 
In the case of PSR B0656+14, {\sl Chandra} observations found evidence for a very faint or compact {\rm PWN} (Birzan et al. 2015, {\emph{submitted}}).
No significant extended emission was seen for $\psr$ with {\sl XMM-Newton} MOS1/2 \citep{deluca2005,Becker2002}. The Continuous Clocking mode observation of $\psr$ by \citet{Pavlov2002} did not allow for exploiting the full spatial resolution capabilities of {\sl Chandra} to search for faint extended emission. Our new {\sl Chandra} ACIS observation, reported here, enables such an investigation.\\
 
We want to note that the distance estimate for $\psr$ has been revised and is different from the value reported in the ATNF catalog\footnote{www.atnf.csiro.au$/$research$/$pulsar$/$psrcat$\/$} \citep{Manchester2005ATNF}.
The dispersion measure of $\psr$ converts to either $d=1.53$\,kpc using the Galactic free electron density model by \citet{Taylor1993}, or to $d=0.73 \pm 0.15$\,kpc using the NE2001 model by \citet{Cordes2002}. \citet{Mignani2010} analyzed optical emission from the pulsar together with its X-ray spectrum. Considering the contributions of the individual thermal components to the multiwavelength spectrum and an upper limit on the neutron star radius of 20\,km, they inferred a much smaller distance, $350\pm 150$\,pc, than the DM-based ones are. 
Thus, $\psr$ is very likely comparably close to us as the other {\sl Musketeers} are. Above all, this distance value also solves the previous puzzle of an unusually high efficiency at $\gamma$-ray wavelengths. Scaling the {\sl Fermi} LAT fluxes reported by \citet{Abdo2010} with the new distance shows that the $\gamma$-ray efficiency of $\psr$ is not so different from other $\gamma$-ray pulsars.

\begin{deluxetable*}{lcccccccccccccc}
\tablecaption{Nearby middle-aged pulsars  \label{pulsars}}
\tablewidth{0pt}
\tablehead{
\colhead{Name} &  \colhead{L/Q\tablenotemark{a}} & \colhead{Age}  & \colhead{$\dot{E}_{33}$\tablenotemark{b}} & \colhead{$B_{11}$\tablenotemark{b}} & \colhead{Dist} & \colhead{Ref\tablenotemark{c}} & \colhead{PM} & \colhead{Ref\tablenotemark{c}} & \colhead{$v_{\rm trans}$} & \colhead{$N_H$} & \colhead{Ref\tablenotemark{c}} & \colhead{$\Gamma_{\gamma}$\tablenotemark{d}}  & \colhead{$E_{\rm cut}$\tablenotemark{d}}  & \colhead{PWN\tablenotemark{e}}\\
 & & \colhead{kyr} & \colhead{} & \colhead{} & \colhead{kpc} &   & \colhead{mas\,yr$^{-1}$} &   & \colhead{km\,s$^{-1}$} & \colhead{$10^{20}$\,cm$^{-2}$} &  &  & \colhead{GeV}}  &  
\startdata
J0357+3205 & RQ & 540 & 6 & 24 & 0.6 & (g,1) & 164 & (6) & 466 & $8\pm 4$ & (13) & $1.0 \pm 0.1$ & $0.8 \pm 0.1$ & Y\\
J0358+5413 & RL & 564 & 45 & 8 & 1 & (p,2) & 12.3 & (7) & 58 & $<72$ & LAB & $\cdots$ & $\cdots$ &     Y\\
J0538+2817 & RL & 618 & 49 & 7 & 1.3 & (p,2) & 57.89 & (8)& 357 & $<54$ & LAB & $\cdots$ &  $\cdots$  &   Y\\
J0622+3749 & RQ & 208 & 27 & 30 & 1.9 & (g,1) & $\cdots$  &	& $\cdots$  & $<21$ & LAB &  $0.6 \pm 0.4$  & $0.6 \pm 0.1$ &   N\\
J0633+1746 & RQ &  342 & 32 & 16 & 0.25 & (p,2) & 169 & (9) & 200 & 1.07 & (14) & $1.2 \pm 0.1$ & $2.2 \pm 0.1$ & Y\\
J0659+1414 & RL & 111 & 38 & 47 & 0.28 & (p,2) & 44.1 & (10) & 59 & $4.3 \pm 0.2$ & (14) & $1.7 \pm 0.5$ & $0.4 \pm 0.2$ & VF\\
J0922+0638 & RL & 497 & 7 & 25 & 1.1 & (p,2) & 88.4 & (11) & 461 & $<3.2$ & LAB & $\cdots$ & $\cdots$ &     N\\
{\bf J1057$-$5226} & RL &  535 & 30 & 11 & 0.35 & (3) & 42 & (3) & 70 & $2.7 \pm 0.2$ & (14) & $1.1 \pm 0.1$ & $1.4 \pm 0.1$ & VF\\
J1741$-$2054 & RL & 386 & 10 & 27 & 0.38 & (p,4) & $109$  & (4)  & $196$  & $12.0^{+0.8}_{-0.7}$  & (4)   & $1.1 \pm 0.1$ & $0.9 \pm 0.1$ &   Y\\
J1745$-$3040 & RL&  546 & 9 & 20 & 0.2 & (p,2) & 7 & (12) & 7 & $<99$ & LAB & $\cdots$ & $\cdots$ &     N\\
J1846+0919 & RQ &  360 & 34 & 15 & 1.5 & (g,1) & $\cdots$  &	& $\cdots$  & $<33$ & LAB & $0.7 \pm 0.3 $ & $2.2\pm 0.5 $ & N\\
J2028+3332 & RQ &  576 & 35 & 9 & 0.9 & (g,5) & $\cdots$  &   & $\cdots$  & $<62$ & LAB & $1.2 \pm 0.2$ &  $1.9 \pm 0.3$  &   N\\
J2030+4415 & RQ & 555 & 22 & 12 & 0.8 & (g,5) & $\cdots$  &	& $\cdots$  & $6^{+15}_{-6}$ & (5) &  $1.6 \pm 0.1$   &  $1.7 \pm 0.3$ &   Y\\
\hline\\
J1530$-$5327 & RL & 944 & 9 & 12 & 1.46 & DM & $\cdots$  &   & $\cdots$  & $<65$ & LAB & $\cdots$ & $\cdots$  &   ?\\
J1549$-$4848 & RL & 324 & 23 & 20 & 1.54 & DM & $\cdots$  &   & $\cdots$  & $<44$ & LAB & $\cdots$  & $\cdots$ &   ?(ND)\\
J1746$-$3239 & RQ & 482 & 33 & 12 & $\cdots$  &   & $\cdots$  &   & $\cdots$  & $<55$ & LAB & $1.4 \pm 0.1$  & $1.5 \pm 0.2$ &   ?\\
J1957+5033 & RQ & 839 & 5 & 17 & 0.9 & (g,1) & $\cdots$  &   & $\cdots$  & $<12$ & LAB &  $1.3 \pm 0.2$  & $1.0 \pm 0.2$ &  ?(ND)
\enddata
\tablenotetext{}{Nearby middle-aged pulsars are defined as having distances $<2$\,kpc, ages between $0.1 -1$\,Myr, and $\dot{E}$ between $5\times 10^{33}- 5\times 10^{34}$\,erg\,s$^{-1}$.}
\tablenotetext{a}{Radio-quiet (RQ) or radio-loud (RL); based on the ATNF catalog (www.atnf.csiro.au$/$research$/$pulsar$/$psrcat$\/$; \citealt{Manchester2005ATNF})}
\tablenotetext{b}{The spin-down energy, $\dot{E}_{33}$, is in units of $10^{33}$ erg\,s$^{-1}$. The magnetic field $B_{11}$ is in units $10^{11}$\,G.}
\tablenotetext{c}{References: (g,1) indicates that the distance was estimated from $\gamma$-ray LAT flux as described by  (1) \citet{Brownsberger2014}; (p,2) denotes parallactic distances corrected for the Lutz-Kelker bias as listed by (2) \citet{Verbiest2012}; (3) \citet{Mignani2010}; DM distance estimate based on the free electron model by \citet{Taylor1993} as listed in the ATNF catalog); (4) \citet{Auchettl2015}; (5) \citet{Marelli2015}; 
(6) \citet{DeLuca2013}; (7) \citet{Chatterjee2004}; (8) \citet{Chatterjee2009}; (9) \citet{Caraveo1996}; (10) \citet{Brisken2003}; (11) \citet{Brisken2003b}; (12) \citet{Zou2005}; (13) \citet{Abdo2013}; (14) \citet{deluca2005};  $N_H$ values with the reference `LAB' indicate upper limits which are the total Galactic H{\sc{I}} column density in the direction of the pulsar using the nh-webtool (heasarc.nasa.gov$/$cgi$-$bin$/$Tools$/$w3nh$/$w3nh.pl) and the LAB survey \citep{Kalberla2005}}
\tablenotetext{d}{The spectrum of each $Fermi$ LAT pulsar was modeled as a power law with an exponential cutoff by \citet{Abdo2013}, their `PLEC1' model. In this table we quote their reported photon
index, $\Gamma_{\gamma}^{\rm PLEC1}$, and  the cutoff energy, $E^{\rm PLEC1}_{\rm cut}$. The cutoff was significantly detected for all pulsars -- all have reported cutoff significance $>33$.}
\tablenotetext{e}{Y -- prominent PWN; N -- no PWN detected in current XMM-$Newton$ or $Chandra$ data; VF -- indication of very faint PWN; ? -- unknown (non-existent X-ray observations with XMM-$Newton$ or $Chandra$); (ND) -- new data expected from scheduled/recent X-ray observations }
\end{deluxetable*}

\section{Observations and Data Analysis}
\label{dataana}
\subsection{{\sl Chandra}}
PSR\,B1055$-$52  was observed with {\sl Chandra} ACIS-I in imaging mode in VFAINT telemetry mode on 2012 October 3rd for 56\,ks (ObsID 13789). We used CIAO (version 4.6.1.) with CALDB (version 4.5.9) for the data reduction and analysis.

We investigated the possibility of pile-up\footnote{Two or more photons are detected as a single event; for more details see cxc.harvard.edu/ciao/ahelp/acis\_pileup.html} since this could influence the spectral analysis and the input for the point spread function (PSF) modeling with MARX\footnote{space.mit.edu/cxc/MARX/index.html}. Using the CIAO tool \texttt{pileup\_map}, we obtained $0.09$\,photons per frame of 3.2\,s in the center of the pulsar image (binned to 1 native ACIS pixel, $0\farcs{49}$). This number translates into a rough estimate of the amount of pileup fraction, 4\,\%, according to the approach by \citet{Davis2001}. 
This amount of pileup is unlikely to significantly influence the pulsar spectrum. Hence, we neglect pile-up for our further data analysis.\\

The data were free of background flares. 
Using \texttt{specextract}, we obtained the pulsar spectrum from a circular region with radius $5\arcsec$. We binned the data with each bin containing at least 25 counts.\\  

In order to evaluate our data for extended emission around the pulsar, we used the ray-trace simulator  MARX \citep{Davis2012} to model the {\sl Chandra} PSF. MARX  (version  5.0) and the associated \textit{Chandra} Ray Tracer (ChaRT; \citealt{Chart2003}) program require an input spectrum for accurate modeling of the pulsar PSF. 
The MARX calibration data are based on CALDB 4.4.7. Since the contamination on the optical-blocking filters of the ACIS detectors\footnote{See e.g., cxc.harvard.edu/ciao/why/acisqecontam.html} is changing the response of the instrument over time, mixing of the CALDB versions in the data and MARX simulations would produce inaccurate results. Therefore we extracted the pulsar spectrum for the MARX modeling from data re-processed with CALDB 4.4.7.
The spectral model files were created from spectral fits using Sherpa and XSPEC.
Following standard CIAO threads, we tested different approaches. In one, the output of ChaRT is used in combination with MARX. In another, the simulation is done solely within MARX. We obtained significantly more counts in the central region of the pulsar using ChaRT$+$MARX than we measured in the actual data. In contrast, the MARX-only simulation produced count numbers comparable to the observed ones. Therefore, we only discuss the latter simulations in this paper. 
In order to minimize the statistical errors of the simulated PSF, we carried out 1000 MARX simulations with an exposure time of 1\,Ms using the internal MARX dither model. 
Our final MARX pulsar PSF model is the average of these 1000 simulations, scaled to the actual exposure time (56\,ks), the considered simulation `errors' are the count errors scaled to the actual exposure time as well. 

\begin{figure}[t!]
\includegraphics[width=8cm]{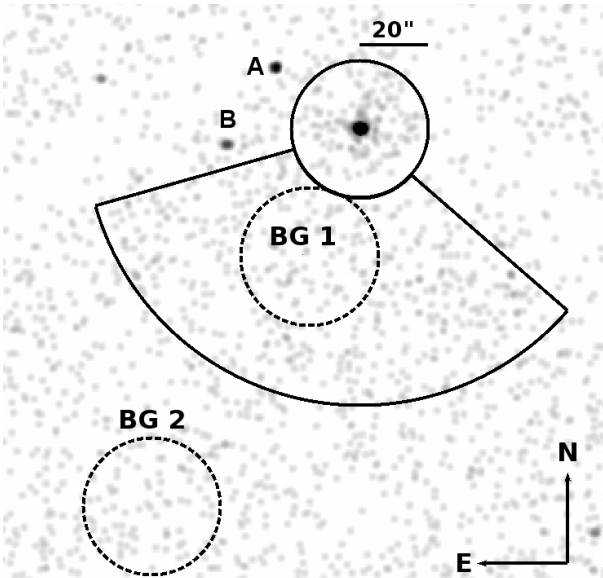}
\caption{Our analysis regions around PSR\,B1055$-$52 which we used to check for large-scale extended emission. The circle on the top marks the maximum radius ($20\arcsec$) we used for annuli around the pulsar. For these measurements background region 1 (BG1) was applied. The fan structure targets larger radii from the pulsar, but avoids contaminating sources A and B, as well as ACIS chip gaps. The image is smoothed and covers the energy range $0.3-8$\,keV.}
\label{PSRWide}
\end{figure} 

\subsection{{\sl XMM-Newton}}
\label{xmmdata}
Our spectral fit results for the {\sl Chandra} pulsar data deviate from previously published results (Section~\ref{respsr}). In order to directly compare our {\sl Chandra} and previous {\sl XMM-Newton} data, we reanalyzed the {\sl XMM-Newton} observations (ObsIDs 0113050101, 0113050201) from 2000 which were presented by \citet{deluca2005}.
The EPIC (European Photon Imaging Camera) observations of $\psr$ employed the pn camera in timing mode, and MOS1 
and MOS2 cameras in imaging mode. For all instruments the medium filter was used. 
We used the {\sl XMM-Newton} Science Analysis Software (SAS), version 13.5 for the re-processing of the observations and data reduction.
These early {\sl XMM-Newton} EPIC-pn observations do not have associated offset maps, and offset map correction or soft photon noise reduction  
is not possible. 
Soft background noise can influence the extracted spectra below 400\,eV \citep{Burwitz2004,deluca2005}, hence the extracted pn spectra only cover energies larger than 400\,eV.

We checked the {\sl Chandra} image for nearby X-ray sources that could contaminate the {\sl XMM-Newton} data (see Figure~\ref{PSRWide}). As noted by \citet{deluca2005}, there is a faint source (source A) in northeast direction. Its angular separation from the pulsar position is $30\arcsec$ in our {\sl Chandra} observation. In the pn timing mode it is $<4$ pixels along the RAWX direction, where spatial information is maintained for this mode \citep{deluca2005}. In Section \ref{respsr}, we use the {\sl Chandra} data to assess the influence of this contaminating X-ray source on the ({\sl XMM-Newton}) pulsar spectral parameters.
There is also a fainter source B at $39\arcsec$ east of the pulsar which is not distinguishable from the broad pulsar PSF in the MOS image. 
There is furthermore a very faint source C in the northeast direction with an angular separation from the pulsar of $3\arcmin$, but it is projected on a similar RAWX pixel as source A in the pn timing observation. Source C is a factor 4 fainter than source A, though.
For the spectral extraction in the pn timing observations we used $33 \le$\,RAWX\,$\le 39$ (both observations 101, 201) for the source regions, and $5 \le$\,RAWX\,$\le 7$ (101) and $4 \le$\,RAWX\,$\le 6$ (201) for the background regions. For the MOS cameras, the source region is a circle centered on the pulsar with radius $45\arcsec$, excluding two circular regions (radius $10\arcsec$) centered on sources A and B. The background spectra were extracted from nearby source-free regions on the same detector chip. 
\begin{figure}[t]
\includegraphics[width=8cm]{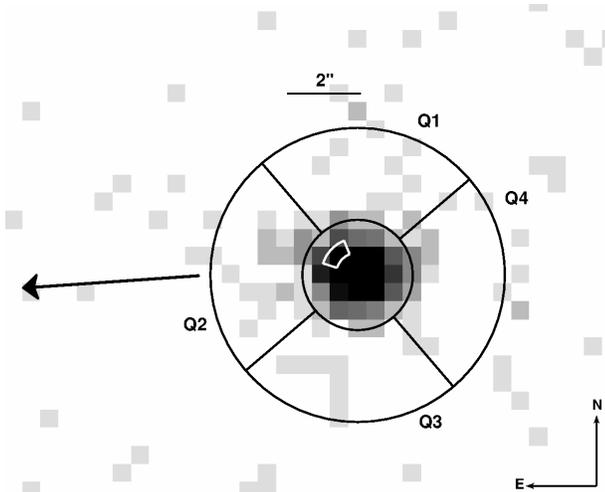}
\caption{PSR\,B1055$-$52 and its close surroundings in X-rays ($0.3-8$\,keV) as seen by {\sl Chandra}. The arrow points in the direction of proper motion 94$^{\circ}\pm$7$^{\circ}$ east of north \citep{Mignani2010}. The annulus region has a size of $1.5\arcsec-4\arcsec$. The four quadrants used for count measurements are labeled. The small wedge inside the annulus indicates the {\sl Chandra} PSF asymmetry region within $1\arcsec$.\vspace{0.2cm} \\}
\label{CloseUpQuad}
\end{figure} 
After filtering for background flares, we extracted spectra using events with pattern 0 to 12 for MOS, but only single and double photon events (pattern 0 to 4) for EPIC-pn. 
The {\sl XMM-Newton} spectra were binned using at least 40 counts per bin.\\

\subsection{Spectral analysis}
\label{specana}
Spectral fits were done using XSPEC version 12.8.1.
We excluded energies higher than 7\,keV since the background dominates at these energies. We used the Tuebingen-Boulder ISM absorption model (\texttt{tbabs}) with the solar abundance table from \citet{Wilms2000}, the photoelectric cross-section table from \citet{Balu1992} together with the He cross-section based on \citet{Yan1998}.
Note that instead of using merged {\sl XMM-Newton} data sets, we obtained simultaneous spectral fits for all the individual six {\sl XMM-Newton} spectra.\\

\section{Results}
\label{results}
In the {\sl Chandra} ACIS data, in a circular region centered on the pulsar with radius $2\arcsec$  and $5\arcsec$ we detected 
$1858 \pm 43$ and $1908 \pm 44$ (background subtracted) counts, respectively, in the energy range $0.3-8$\,keV.
There is no obvious large scale extended emission (Figure~\ref{PSRWide}) but a slight asymmetry of the count distribution within $4\arcsec$ of the pulsar (Figure~\ref{CloseUpQuad}). 

\subsection{Extended emission around the pulsar}
\label{resextended}
We consider three different energy bands for our extended emission analysis: the broad energy band ($0.3-8$\,keV, BEB), the soft energy band ($0.3-1$\,keV, SEB), and the hard energy band ($1-8$\,keV, HEB). As explained in Section~\ref{dataana}, we employ for comparison MARX simulations for 1\,Ms, scaled to exposure time of the observation. We tested different `AspectBlur' parameter values and found $0\farcs{2}$ to best reproduce the radial count distribution of the inner PSF of the observed pulsar. 

\subsubsection{Radial enhancement of surface brightness}
\label{radial}
\begin{figure}
\includegraphics[width=8.5cm]{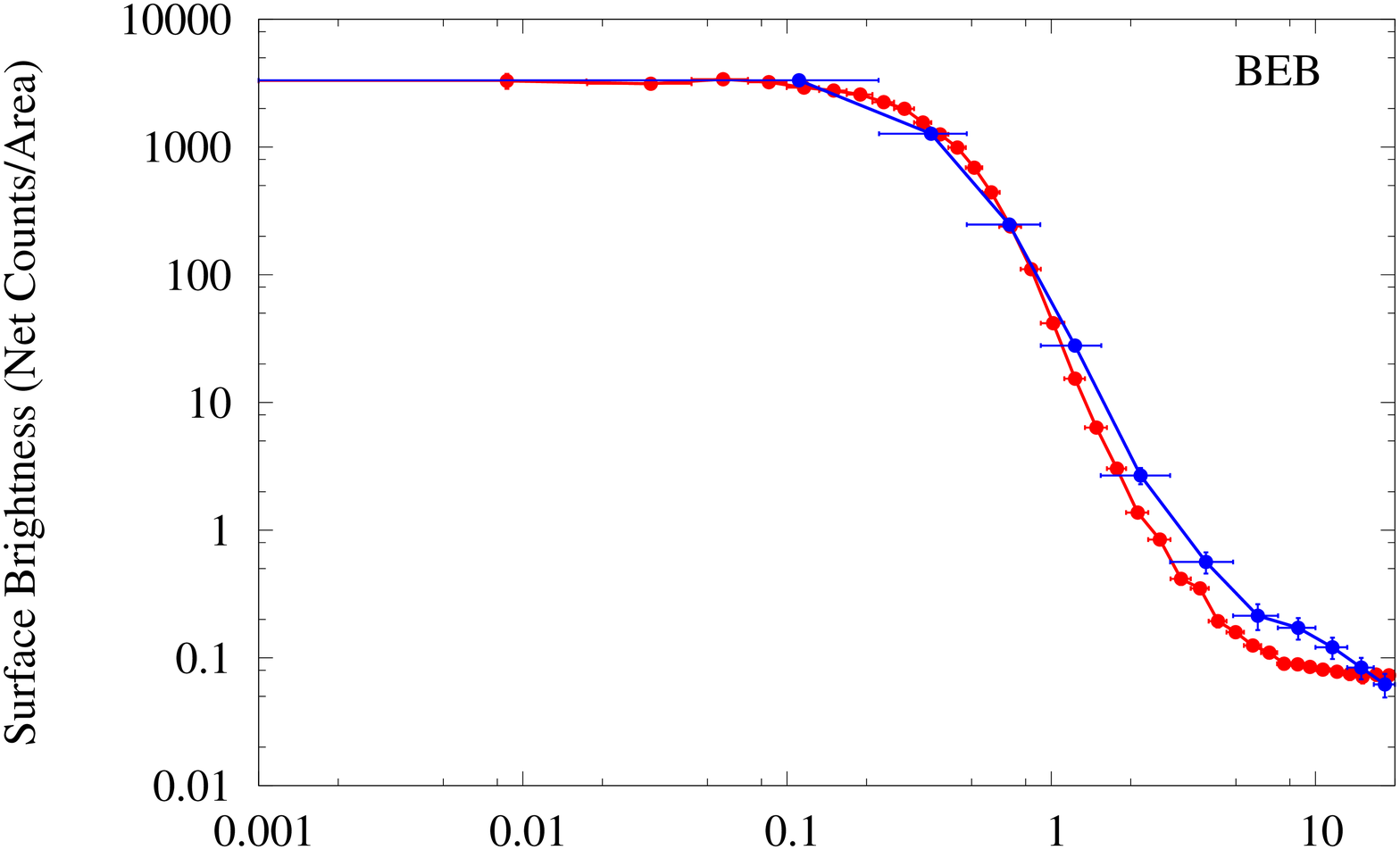}
\includegraphics[width=8.5cm]{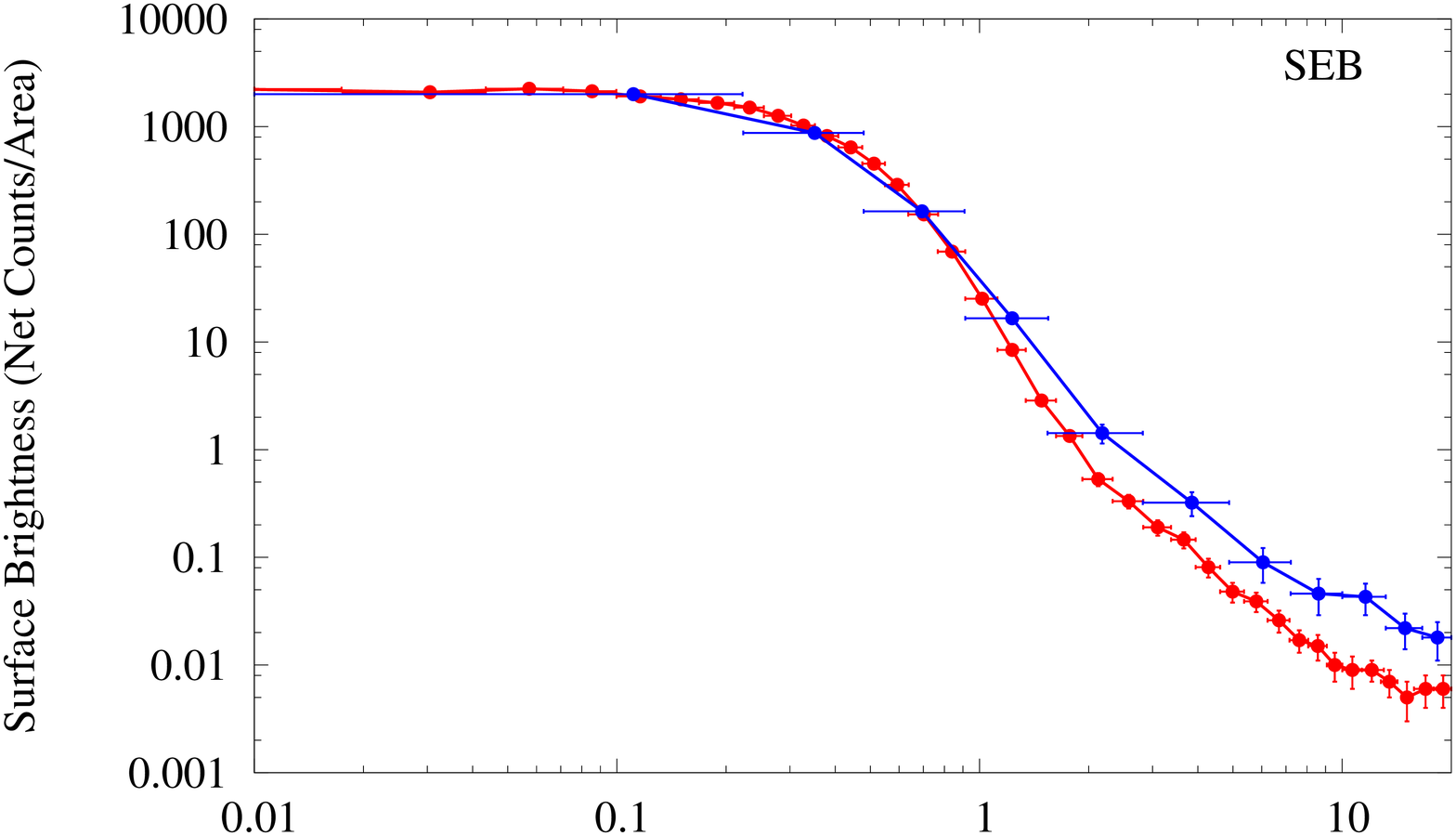}
\includegraphics[width=8.5cm]{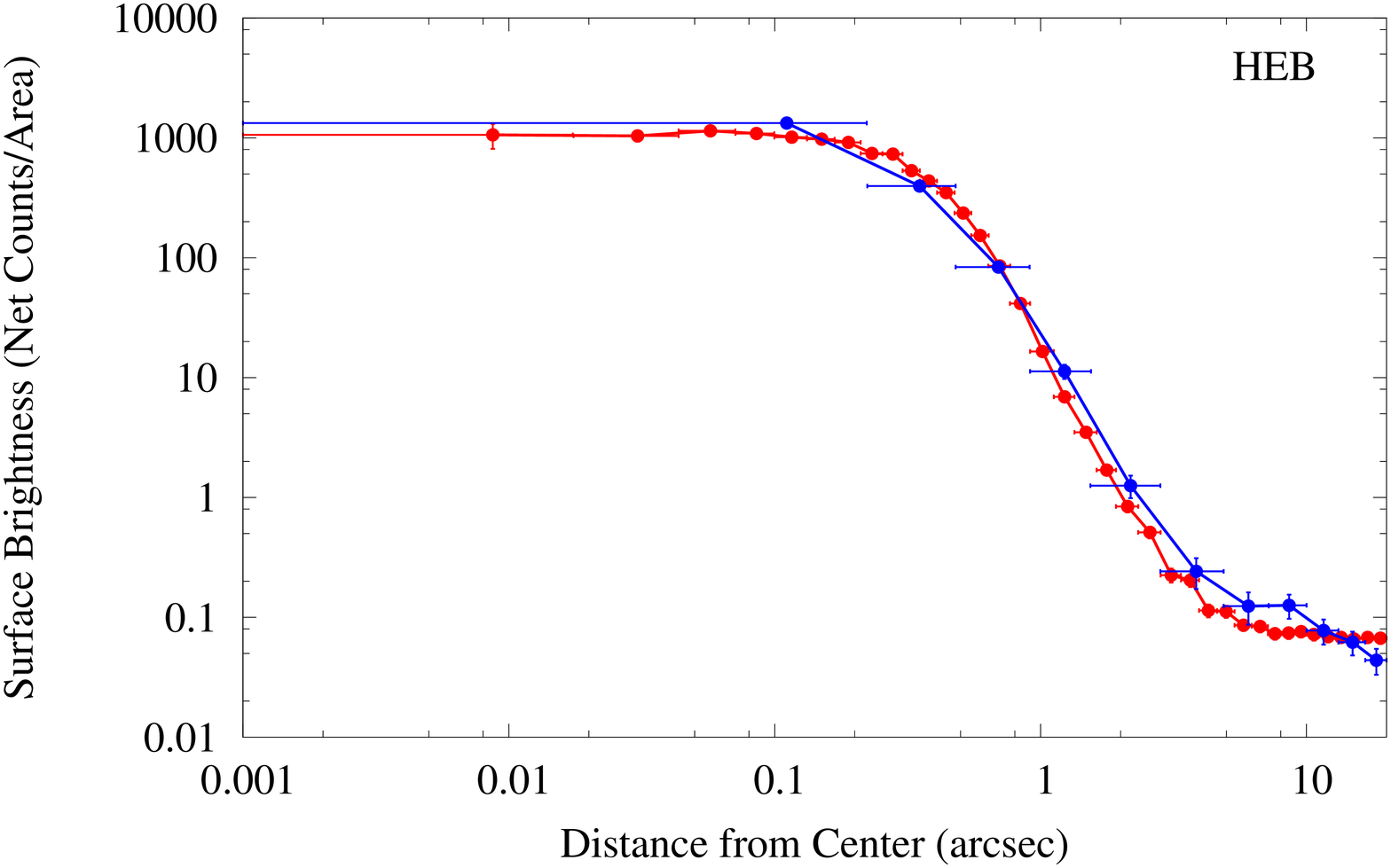}
\caption{The radial surface brightness distribution in the observation (blue) and the MARX simulation (red; 1\,Ms exposure time, scaled to source exposure time, internal dither model, AspectBlur$=0\farcs{2}$) in the broad energy band (top panel), the soft energy band (middle panel), and the hard energy band (bottom panel). The `error' bars in distance direction indicates the bin size used for the surface brightness  measurements, the error bar in $y$ direction is the actual uncertainty for the respective bin. Note that no background subtraction was done for the observational data. Instead, the respective background level was added to the MARX simulation surface brightness. }
\label{MARX}
\end{figure}

The radial distributions of the surface brightness of the MARX simulation and the observation are shown in Figure~\ref{MARX}. For this plot, the background is not subtracted from the observational data. Instead, the respective background surface brightness of the observation was added to the MARX simulation. The surface brightness of the background is 0.071, 0.005, and 0.066\,counts\,arcsec$^{-2}$ for the BEB, SEB and HEB, respectively.
The uncertainty level\footnote{The error for the simulated surface brightness, $n_{\rm sim}$, was calculated as: $\delta n_{\rm sim}=A^{-1}(N^{\rm sc}_{\rm sim} t_{\rm exp}/t_{\rm sim} + N_{\rm BG} A^{2}/A^{2}_{\rm BG})^{1/2}$,
where $t_{\rm exp}$ is the exposure time, $t_{\rm sim}$ is the simulation exposure time, $N^{\rm sc}_{\rm sim}$ is the number of counts scaled by the exposure time, $N_{\rm BG}$ is the number of background counts, $A$ is the area of the respective annulus, and $A_{\rm BG}$ is the area of the background region.} of the MARX simulation curve at the largest distances is actually dominated by the uncertainty of the added observational background surface brightness.\\

\begin{deluxetable}{lcc|cc|cc}[t]
\tablecaption{Significance of the radial flux enhancement. \label{Gap}}
\tablewidth{0pt}
\tablehead{
\colhead{} & \multicolumn{2} {c} {$0.3-8.0$\,keV} & \multicolumn{2} {c} {$0.3-1.0$\,keV} & \multicolumn{2} {c} {$1-8.0$\,keV}\\
\colhead{} & \colhead{Obs} & \colhead{Sim} & \colhead{Obs} & \colhead{Sim} & \colhead{Obs} &\colhead{Sim} }\\
\startdata
$r_{\rm in}$ [arcsec] & 4.88 & 4.61 & 4.88 & 4.61 & 4.88 & 4.61 \\
$r_{\rm out}$ [arcsec] & 13.19 & 14.21 & 20.00 & 20.00 & 10.01 & 10.01 \\
$N$ \tablenotemark{(a)} & 73.00 & 10.30 & 39.00 & 6.08 & 30.00 & 3.55 \\
$\delta N$ & 8.54 & 3.21 & 6.24 & 2.47 & 5.48 & 1.88 \\
$A$ [arcsec$^2$] & 472 & 568 & 1182 & 1190 & 240 & 248 \\
$n$ \tablenotemark{(b)} & 0.16 & 0.089 &  0.033 & 0.010 & 0.13 & 0.081 \\
$\delta n$ & 0.02 & 0.008 & 0.005 & 0.002 & 0.02 & 0.008 \\
$n_{\rm obs}-n_{\rm sim}$ & \multicolumn{2} {c} {0.066} & \multicolumn{2} {|c} {0.0231} & \multicolumn{2} {|c} {0.044} \\
$\delta (n_{\rm obs}-n_{\rm sim})$ & \multicolumn{2}{c}{0.020} & \multicolumn{2}{|c}{0.0057} & \multicolumn{2}{|c}{0.024} \\
$\mathcal{S}$\tablenotemark{(c)} [$\sigma$] & \multicolumn{2}{c}{3.3} & \multicolumn{2}{|c}{4.1} & \multicolumn{2}{|c}{1.8} 
\enddata
\tablenotetext{(a)}{Source counts for the observation (``Obs'') and scaled counts of the 1\,Ms MARX simulation (``Sim'')} 
\tablenotetext{(b)}{Surface brightness [cts\,arcsec$^{-2}$] includes background counts for the observational data. The background surface brightness is added to the scaled surface brightness of the 1\,Ms MARX simulation} 
\tablenotetext{(c)}{Surface brightnesses, $n$, their differences, uncertainties, and the significance, $\mathcal{S}$, were calculated before rounding the respective contributing values.} 
\end{deluxetable}

There is an enhancement of X-ray emission starting from about $2\arcsec$ to about $20\arcsec$. It is especially well visible in the soft band.
We evaluated the significance of the enhancement in Table~\ref{Gap}. Since the MARX simulation allowed a finer binning, the considered observation and simulation annuli do not have exactly the same size. For the comparison in Table~\ref{Gap}, we used for the simulation annuli with areas at least as large or larger than the areas of the annuli utilized for the observational data. This ensures that our significance estimate in Table~\ref{Gap} represents a lower limit to the actual significance of the enhancement in the considered region.  
The enhancement in the surface brightness has a significance of $3\sigma$ in the BEB and $4\sigma$ in the SEB, while in the HEB the surface brightness in the observation is consistent with the simulation.  
We checked that the enhancement persisted even if we used different (and badly fitting) `AspectBlur' values (the probed range was $0\farcs{07} - 0\farcs{8}$). The reason is that only the inner part (within $\approx 4\arcsec$) of the PSF is significantly changed by this MARX parameter.\\ 

In order to crudely estimate the flux in the X-ray emission enhancement in the BEB, we applied the \texttt{eff2evt} tool which uses the event energies and effective areas at the event locations for the flux estimate.  
We used the $4.88\arcsec-13.19\arcsec$ annulus (as in Table~\ref{Gap}) for the data and the MARX simulation, and did the same for the (area-scaled) background region. The resulting flux for the excess emission is $1.0 \pm 0.2 \times 10^{-14}$\,erg\,s$^{-1}$\,cm$^{-2}$ for the BEB ($0.3-8$\,keV). 
Alternatively, using the count rate of the BEB enhancement in the same annulus ($5.5 \times 10^{-4}$\,cps), PIMMS\footnote{cxc.harvard.edu/toolkit/pimms.jsp} with {\sl Chandra}'s AO13 response, and an absorbed ($N_{\rm H}=3.4\times 10^{20}$\,cm$^{-2}$ from the XMM/{\sl Chandra} fit of the pulsar spectrum in Table~\ref{specfit}) power law with photon index range of $\Gamma=2$ and $\Gamma=4$, we obtain absorbed (unabsorbed) BEB fluxes of $0.6 \times 10^{-14}$ ($0.7 \times 10^{-14})$\,erg\,s$^{-1}$\,cm$^{-2}$ and $1.1 \times 10^{-14}$ ($1.7 \times 10^{-14}$)\,erg\,s$^{-1}$\,cm$^{-2}$.\\

We checked whether there is any flux enhancement at angular separations $20\arcsec$ to $60\arcsec$. For this we use fan-shaped regions to avoid chip gaps. Although we used a different background region (see Figure~\ref{PSRWide}), the background rates from this region are consistent with the ones quoted above for the BEB, SEB, and HEB. We did not find any significant flux enhancement for angular separations $>20\arcsec$.

\subsubsection{Probing for asymmetry in the extended emission}
\label{extasymmetry}
\begin{figure}[b]
\includegraphics[width=8.5cm]{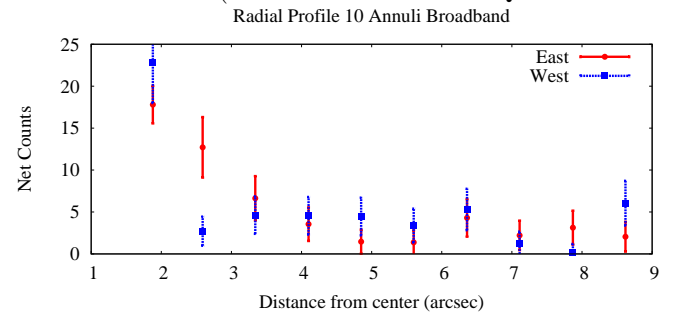} 
\caption{Counts in the East and West direction over distance from the pulsar in the BEB. The measurement annuli start at $1\farcs{5}$.}
\label{GraphNetCountsBB10}
\end{figure}
Figure~\ref{CloseUpQuad} appears to show a slight asymmetry of the count distribution within $4\arcsec$ of the pulsar. This asymmetry is notably in the direction of the $\psr$ proper motion.
Aiming to investigate this asymmetry, we compare in Figure~\ref{GraphNetCountsBB10} the East and West side of the close neighborhood of the pulsar ($\ge1\farcs{5}$ to exclude the known PSF asymmetry region\footnote{see cxc.harvard.edu/ciao4.6/caveats/psf\_artifact.html}). There are more counts in the second annulus (center at $2\farcs{6}$) for the East data than for the West data. Count numbers for these 10 annuli are, however, so small that $2\sigma$ errors overlap for the annulus in question. Thus, the difference between East and West is not statistically significant.\\

Based on our MARX simulation, we also tried another approach. We divided the immediate surrounding of the pulsar in four annular ($1\farcs{5}-4\arcsec$) quadrants (Q1: 319$^{\circ}-49^{\circ}$, where $0^{\circ}$ and $90^{\circ}$ indicate the North and East direction, respectively, Q2: 49$^{\circ}-139^{\circ}$, Q3: 139$^{\circ}-229^{\circ}$, Q4: 229$^{\circ}-319^{\circ}$; see Figure~\ref{CloseUpQuad}).
Similarly to the annuli estimates in Table~\ref{Gap} in Section~\ref{radial}, we calculated the difference of the surface brightness considering the observed and the simulated MARX data in the BEB, SEB, and HEB in each quadrant. The observed respective background surface brightness of the observation was added to the simulation data for the comparison. 
While we indeed find a slight enhancement in the second quadrant as expected from Figure~\ref{CloseUpQuad}, the significance is just about $3\sigma$ for the BEB and SEB for the available small number of counts (24 counts maximum in Q2 of which 10 are expected to be pulsar and background counts. Thus, we cannot firmly exclude that the hint of asymmetry for the emission around the pulsar in the $1\farcs{5}-4\arcsec$ region is just a statistical fluctuation in the spatial count distribution.

\begin{deluxetable}{lcccc}[t]
\tablecaption{Significance of extended emission in quadrants.\label{quadrants}}
\tablewidth{0pt}
\tablehead{
\colhead{} & \colhead{Q1} & \colhead{Q2} & \colhead{Q3} & \colhead{Q4}}\\
\startdata
BEB $n_{\rm obs}$ & 1.30$\pm$0.35 & 2.22$\pm$0.45 & 1.02$\pm$0.31 & 1.39$\pm$0.36 \\
BEB $n_{\rm sim}$\tablenotemark{(a)} & 1.00$\pm$0.07 & 0.90$\pm$0.07 & 0.73$\pm$0.06 & 1.0$\pm$0.07 \\
BEB $\mathcal{S}$\tablenotemark{(b)} & 0.8           & 2.9 & 0.9 & 1.1 \\
SEB $n_{\rm obs}$ & 0.28$\pm$0.16 & 1.30$\pm$0.35 & 0.65$\pm$0.25 & 1.02$\pm$0.31 \\
SEB $n_{\rm sim}$\tablenotemark{(a)} & 0.44$\pm$0.05 & 0.38$\pm$0.04 & 0.30$\pm$0.04 & 0.39$\pm$ 0.05 \\
SEB $\mathcal{S}$\tablenotemark{(a)}  & 1.0 & 2.6 & 1.4 & 2.0 \\
HEB $n_{\rm obs}$ & 1.01$\pm$0.31 & 0.93$\pm$0.29 & 0.37$\pm$0.19 & 0.37$\pm$0.19 \\
HEB $n_{\rm sim}$\tablenotemark{(a)} & 0.56$\pm$0.05 & 0.52$\pm$0.05 & 0.43$\pm$0.04 & 0.60$\pm$0.05 \\
HEB $\mathcal{S}$\tablenotemark{(b)}  & 1.4 & 1.4 & 0.3 & 1.2 
\enddata
\tablenotetext{(a)}{The surface brightness of the simulation includes the observational background component as described in the text.} \tablenotetext{(b)}{Significances, $\mathcal{S}$, of the surface brightness difference were calculated before rounding of the surface brightness values.} 
\end{deluxetable}

\subsection{The pulsar}
\label{respsr}
\begin{deluxetable*}{lc|c|cc|c}
\tablecaption{Spectral fit parameters of the pulsar for different instruments.\tablenotemark{a} \label{specfit}}
\tablewidth{0pt}
\tablehead{
\colhead{Parameter\tablenotemark{b}} & \colhead{pn Only} & \colhead{pn\&MOS} & \colhead{pn\&MOS\tablenotemark{c}} & \colhead{ACIS\tablenotemark{c}} &  \colhead{ACIS Only}}\\
\startdata
$N_{\rm H}$ (10$^{19}$ cm$^{-2})$ & 88 $\pm$ 5 & 33 $\pm$ 1 & 34 $\pm$ 1 & 34 (tied) & 240$^{+200}_{-150}$ \\
$kT_{\rm CBB}$ (eV) & 61 $\pm$ 5 & 69$^{+2}_{-3}$ & 68 $^{+2}_{-3}$ & 81 $\pm$ 5 & 63$^{+12}_{-10}$ \\
$R_{\rm CBB}$ (km) & 10$^{+3.8}_{-3.0}$ & 5.3 $\pm$ 0.6 & 5.6$^{+0.6}_{-0.7}$ & 2.8$^{+0.5}_{-0.4}$ & 14$^{+53}_{-13}$ \\
$kT_{\rm HBB}$ (eV) & 140 $\pm$ 20 & 160$^{+20}_{-10}$ & 160$^{+20}_{-10}$ & 220$^{+50}_{-40}$ & 170$^{+50}_{-40}$ \\
$R_{\rm HBB}$ (m) & 320$^{+140}_{-110}$ & 190$^{+57}_{-31}$ & 190$^{+57}_{-33}$ & 67$^{+35}_{-27}$ & 180$^{+230}_{-100}$ \\
$\Gamma$ & 1.9 $\pm$ 0.4 & 1.9 $\pm$ 0.3 & 1.9 (tied) & 1.9 $\pm$ 0.2 & 2.3$^{+0.5}_{-0.6}$ \\
$\mathcal{N}_{\rm PL}$ & 2.4$^{+1.1}_{-0.8}$ & 2.0$^{+0.6}_{-0.5}$ & 2.1$^{+0.6}_{-0.5}$ & 1.7$^{+0.6}_{-0.7}$ & 2.9$^{+2.3}_{-1.7}$ \\
$f_{\rm abs, 0.3-8 keV}$ EPIC\tablenotemark{d}  & 137$^{+39}_{-6}$ & 152$^{+5}_{-2}$ & 152$^{+6}_{-4}$ & $\cdots$ & $\cdots$ \\
$f_{\rm abs, 0.3-8 keV}$ ACIS\tablenotemark{d}  & $\cdots$ & $\cdots$ & $\cdots$ & 108$^{+10}_{-9}$ & 80$^{+21}_{-11}$ \\

$f^{\rm PL}_{\rm unabs, 0.5-8 keV}$\tablenotemark{d} & 12 $\pm$ 2 & 9.9$^{+1.1}_{-0.2}$ & 10 $\pm$ 1 & 8.3$^{+2.1}_{-1.8}$ & 11$^{+5}_{-4}$ \\

$L^{\rm PL}_{\rm 0.5-8 keV}$ (10$^{30}$ erg s$^{-1}$) & 1.73 $\pm$ 0.25 & 1.38$^{+0.16}_{-0.03}$ & 1.47 $\pm$ 0.16 & 1.22$^{+0.31}_{-0.27}$ & 1.63$^{+0.76}_{-0.54}$ \\
$L_{\rm bol, CBB}$ (10$^{31}$ erg s$^{-1}$) & 19.3$^{+12.7}_{-9.1}$ & 8.4$^{+1.7}_{-1.2}$ & 8.6$^{+1.7}_{-1.4}$ & 4.3$^{+1.0}_{-0.7}$ & 39.9$^{+292}_{-32}$ \\
$L_{\rm bol, HBB}$ (10$^{30}$ erg s$^{-1}$) & 4.9$^{+3.4}_{-1.9}$ & 3.2$^{+1.0}_{-0.6}$ & 3.2$^{+1.0}_{-0.7}$ & 1.4$^{+0.7}_{-0.5}$ & 3.6$^{+8.1}_{-2.1}$ \\
$\chi^{2}$/dof & 0.87 & 0.99 & 0.96 & 0.96 & 0.79 \\
dof & 87 & 221 & 274 & 274 & 52 \\
$\mathcal{F}_{\rm pn}$ & $\cdots$ & 1 (fixed) & 1 (fixed) & $\cdots$ & $\cdots$ \\
$\mathcal{F}_{\rm MOS1}$ & $\cdots$ & 1.06 $\pm$ 0.02 & 1.06 $\pm$ 0.02 & $\cdots$ & $\cdots$ \\
$\mathcal{F}_{\rm MOS2}$ & $\cdots$ & 1.04 $\pm$ 0.02 & 1.04 $\pm$ 0.02 & $\cdots$ & $\cdots$ \\
$\mathcal{F}_{\rm ACIS}$ & $\cdots$ & $\cdots$ &$\cdots$ & 1 (fixed) &$\cdots$ \\
\enddata
\tablenotetext{a}{A model of two BBs and a PL was used. Uncertainties indicate the 90\% confidence level. Luminosity uncertainties do not include the distance uncertainties. Luminosity and radius calculations were done using a distance value of 350\,pc and the exact best-fit parameters before rounding.}
\tablenotetext{b}{The following best-fit parameters are given: absorbing hydrogen column density $N_{\rm H}$, BB temperature $kT$, normalization $\mathcal{N}_{\rm PL}$ of the PL component in units of 10$^{-5}$\,photons\,keV$^{-1}$\,cm$^{-2}$\,s$^{-1}$ at 1 keV, radius $R$ for the equivalent sphere of the BB emission, photon index $\Gamma$ of the PL component, fluxes $f$, bolometric luminosities of the BBs, the luminosity of the PL component in the energy range $0.5-8$\,keV, reduced $\chi^2$, the number of degrees of freedom, and the calibration factors $\mathcal{F}$ in the case of tied fits,}
\tablenotetext{c}{pn, MOS1, MOS2, and ACIS were fitted together. The $N_{\rm H}$ of {\sl Chandra} was tied to the {\sl XMM} N$_{H}$, and the {\sl XMM} photon index was tied to the {\sl Chandra} photon index.}
\tablenotetext{d}{Fluxes are given in units of 10$^{-14}$ erg cm$^{-2}$ s$^{-1}$.\vspace{0.8cm}} 
\end{deluxetable*}

Details on the $\psr$ spectrum were previously reported for the sensitive \emph{XMM-Newton} observations ($\approx 120,500$\,counts for all EPIC detectors; \citealt{deluca2005}).
Our ACIS pulsar spectrum has much fewer counts ($\approx 1900$). Fitting the data with a similar spectral model as used by \citet{deluca2005}, two blackbodies (BBs) and a power law (PL) with interstellar absorption, we obtained an acceptable fit to our ACIS spectrum (last column in Table~\ref{specfit}). We found differences with the values reported by \citet{deluca2005}. The difference in the absorbing column density is a factor 10, which cannot be explained by different abundances or our large uncertainties. We also found an indication for a larger (i.e., softer) PL index, although our uncertainty range and the one of \citet{deluca2005} still overlap for this parameter.
In contrast to \emph{XMM-Newton}, \emph{Chandra} is able to clearly resolve neighboring sources. In particular, we find not only source A, but also an additional fainter source B located within the MOS PSF (Sources A and B are labeled in Figure~\ref{PSRWide}). 
While \citet{deluca2005} exclude regions encompassing the source A from the MOS data, this source cannot be completely excluded from the pn timing data.
It is interesting to check the influence source A would have on the pulsar spectrum if it were included in the spectral extraction region.
As \citet{deluca2005} noted, Source A is much fainter than the pulsar, and it is a hard source. In our ACIS data, Source A has 77 counts in comparison to the pulsar which has 1912 (using circular extraction regions with radii of $5\arcsec$ for both).  
Using `cstat' statistics and a simple PL, we find a photon index  $\Gamma=1.5^{+1.0}_{-0.7}$ for Source A.
If we fit the pulsar spectrum including the counts of Source A with the 2BB$+$PL model, we find a PL index 
$\Gamma=2.0^{+0.5}_{-0.7}$ versus $\Gamma=2.3^{+0.5}_{-0.6}$ if only the pulsar spectrum is fit. 
While this is not a significant change, the numbers seem to support the natural expectation for this hard source that the PL index would only get smaller if Source A cannot be properly removed in the spectral extraction region. We, however, found the opposite -- the (Source A-free) ACIS spectral model fit gave a larger photon index than what was found with \emph{XMM-Newton} by \citet{deluca2005}.\\ 

Guided by the new spatial information from \emph{Chandra} we reanalyzed the \emph{XMM-Newton} data as described in Section~\ref{xmmdata}.  
Due to our more rigorous source extraction regions, we have fewer counts than \citet{deluca2005}. Overall, we obtained for our source extraction regions 30,644 (source and background) counts, 10,965 counts, and 11,734 counts for pn, MOS1 and MOS2, respectively. 
As noted in Section~\ref{specana}, we did simultaneous spectral fits for all the individual six {\sl XMM-Newton} spectra from the two respective observations.
In Table~\ref{specfit}, we present our fit results for the 2BB+PL model considering only the pn data (second column) and a tied fit for the pn and MOS data.
While there are differences in these two fits, they are usually not significant and well within cross-calibration errors of the \emph{XMM-Newton} instruments \citep{Read2014}. 
The only exception is $N_H$ where the pn-only fit gave a significantly higher value. As explained in Section~\ref{xmmdata}, these early pn timing observations are a rare case where there are no usable counts below 400\,eV in pn, but some counts for MOS. Since interstellar absorption has its largest effect on soft X-rays, $N_H$ from the tied fit is more reliable.
We find a slight shift to a softer PL photon index in comparison to the results by \citet{deluca2005} ($1.9\pm 0.3$ versus $1.7 \pm 0.1$). Reasons for this small change are our more rigorous extraction regions, but probably also calibration updates for the MOS instruments.\\ 

\begin{figure}[h!]
\includegraphics[height=8.5cm, angle=-90]{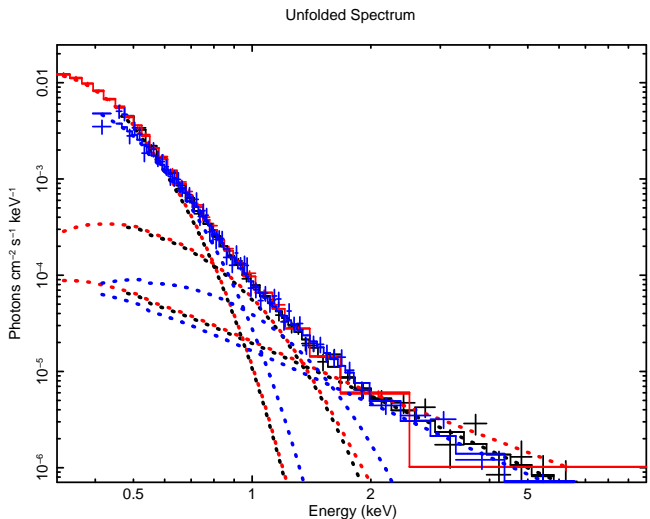} 
\caption{The unfolded spectrum of the pulsar for the pn (black), MOS1 and MOS2 (combined, red) and ACIS (blue) instruments using the spectral model fit where $N_H$ and the PL photon indices are tied for the {\em XMM-Newton} and {\em{Chandra}} data (columns 4 and 5 in Table~\ref{specfit}). Note that we do a rigorous simultaneous fit of all individual 7 spectra, but for better visibility, we show here the combined events of the pn, MOS and ACIS instruments, respectively.}
\label{xmmcxouf}
\end{figure}

\begin{figure}[t]
\includegraphics[width=8cm]{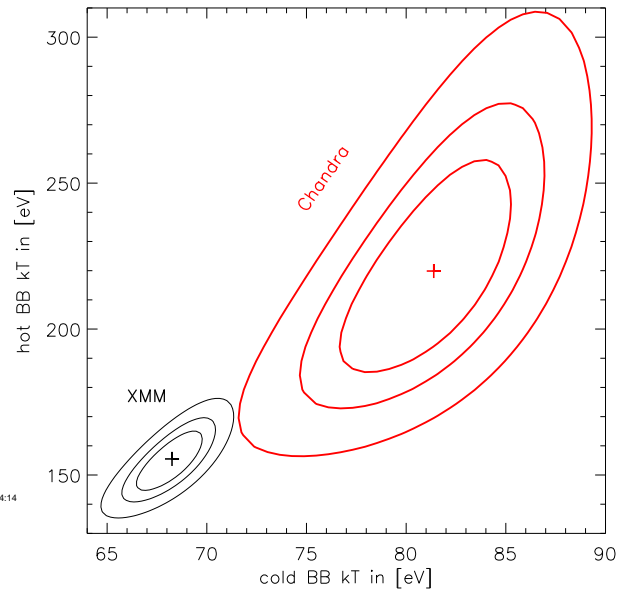} 
\caption{Blackbody temperature confidence contours (68\,\%, 90\,\%, 99\,\%) for the combined {\em XMM-Newton}/{\em{Chandra}} fit ($N_H$ and the PL photon indices are tied, see columns 4 and 5 in Table~\ref{specfit}). The {\em XMM-Newton} contours are black, {\em{Chandra}} contours are red.}
\label{CBBvHBB}
\end{figure}

We fit the 2BB$+$PL model to the \emph{XMM-Newton} data in combination with our ACIS data using tied $N_H$ values (since ACIS is not very sensitive at the lowest energies) and tied photon indices (assuming ACIS might give a more realistic value because of better spatial resolution and lower background at higher energies). 
The results for these tied parameters, listed in column 4 and 5 in Table~\ref{specfit}, are very close to the \emph{XMM-Newton}-only fit (third column) as one would expect considering the different count statistics.
An unfolded spectrum for the resulting best-fit model is shown in Figure~\ref{xmmcxouf}.
It is interesting to note that even with these two parameters tied, there are significant differences between ACIS and the \emph{XMM-Newton} fit parameters. This is true for the BB emission areas, the cold BB temperature, but also for the observed flux. The observed ACIS flux in the energy range $0.3-8$\,keV is only 71\% of the \emph{XMM-Newton} EPIC flux in the same energy range. 
In Figure~\ref{CBBvHBB}, we show contour levels for the BB temperatures. 
Interestingly, even the 99\% temperature contours do not overlap.  We discuss possible reasons in Section~\ref{psrdiscuss}.\\

As \citet{deluca2005} have already showed, the utilization of the BB model as description for the thermal emission of $\psr$ requires at least two regions with different temperatures and areas and an additional PL component. In order to check if the spectrum of $\psr$ can be fitted with a two-component (thermal$+$PL, or thermal$+$thermal) model, we employed NS atmosphere models for the thermal emission.  
Since the work by \citet{deluca2005}, new NS atmosphere models have become available, e.g., the NS Magnetic Atmosphere Models (NSMAXG) by \citet{Mori2007,Ho2008,Ho2013} as implemented in XSPEC.
We used the pn data to carry out fits using a NSMAXG$+$PL for a magnetic field $B=10^{12}$\,G, fixed NS mass of $1.4$\,M$_{\odot}$, NS radius of 10\,km and distance of 0.35\,kpc.
We checked fits with a hydrogen atmosphere (specfiles 1200, 123100, 123190 in XSPEC), carbon atmosphere (specfile 12006), oxygen atmosphere (specfile 12008), neon atmosphere (specfile 12010). The fits were generally not acceptable (reduced $\chi^2>2$ for 89 dof) even if we allowed the mass, radius and distance to vary.
We obtained the best fit  for the hydrogen atmosphere (specfile 1200). However, the reduced $\chi^2=1.4$ for 89 dof is significantly worse than the one of the 2BB$+$PL fit ($\chi^2=0.9$ for 87 dof), and there are substantial systematics in the soft part of the spectrum. Furthermore, the obtained normalization, $\mathcal{N}=R^2_{\rm Em}/R^2_{\rm NS}=34$ (where $R_{\rm Em}$ indicates the size of the emission region), is rather unrealistic. Instead of a NSMAXG$+$PL we also tried NSMAXG$+$NSMAXG for this hydrogen atmosphere to consider possible hot spots, but the fit was not accetable ($\chi^2=1.6$ for 89 dof, systematic residuals) either. 
We further checked a three-component model, NSMAXG$+$NSMAXG$+$PL for the hydrogen atmosphere (specfile 1200), but the fit was not accetable ($\chi^2=1.3$ for 87 dof, systematic residuals, $\mathcal{N}=R^2_{\rm Em}/R^2_{\rm NS}=395^{+6}_{-138}$) either. 
Overall, for a two- or three-component model approach, the NSMAXG (hydrogen) models seem not to be applicable to \psr.\\

\section{Discussion}
\subsection{Extended Emission}
\label{disext}
We detected extended emission around $\psr$ with a significance of $4\sigma$ (Table~\ref{Gap}) in the  SEB for a region $4\farcs{9}-20\arcsec$. 
In Figure~\ref{MARX}, the presence of the radial surface brightness enhancement of the observational data with respect to the MARX simulation is already visible at smaller radii, starting from $1\farcs{5}$. Based on Figure~\ref{MARX}, we did not use these smaller radii in order to optimize the significance analysis. However, the investigation of the $1\farcs{5}-4\arcsec$ region around the pulsar  (Figure~\ref{CloseUpQuad}, Table~\ref{quadrants}) resulted in finding of a $3\sigma$ asymmetry in the present emission (there is no significant asymmetry at larger radii), with most counts in the SEB as well.\\ 

It is difficult to derive further spectral and spatial information for the extended emission from this ACIS observation, because even in the BEB there are only $\approx 30$ (background-subtracted) counts for the chosen extraction region listed in Table~\ref{Gap}. Since the extended X-ray emission at $4\farcs{9}-20\arcsec$ is only significant in the SEB, a possible explanation for the excess could be a dust scattering halo, whose spectrum is generally softer than the central source.
So far, no such halos have been reported for any of the neutron stars which have a similar low interstellar absorption ($N_H \approx 10^{20}$\,cm$^{-2}$).\\
 
As described in Section~\ref{resextended}, we did not find any significant flux enhancement for angular separations $>20\arcsec$. 
Since detected dust scattering halo profiles are usually very broad (see, e.g., \citealt{Predehl1995}), it is interesting to check whether the surface brightness profile around $\psr$ is consistent with a dust scattering halo profile.
Based on the works by \citet{Draine2003} and  \citet{Predehl1995}, we use the following formula for the halo flux, $F_{\rm halo} (\theta_1,\theta_2, E)$, in an annulus $\theta_1 < \theta_h < \theta_2$ at a photon energy $E$:
\begin{equation}
F_{\rm halo} (\theta_1,\theta_2, E) =  F_{\rm PSR} (E) \tau_{\rm sca} (1\,{\rm keV}) \frac{\pi (\theta_2-\theta_1)}{2 \Theta E},
\label{haloflux}
\end{equation}
where $F_{\rm PSR}$ is the flux of the pulsar, $\tau_{\rm sca} (1\,{\rm keV})$ is the scattering optical width at 1\,keV, and the constant $\Theta$ depends on the dust model (we use $\Theta =360\arcsec$ as derived by \citealt{Draine2003} for a dust model by \citealt{Weingartner2001}). Formula (\ref{haloflux}) is only applicable for small scattering angles, $\theta_2 \ll \Theta$.
In order to use the measured count numbers instead of fluxes, one has to convolve the respective fluxes with the instrument response. 
The transformed formula for count numbers is very similar to equation~(\ref{haloflux}) (for details, see, e.g., B\^irzan et al. 2015, \emph{submitted}), but instead of $(1/E)$ uses $(1/E)_{\rm eff}$, the mean inverse energy of the central source photons. For $\psr$, $(1/E)_{\rm eff}=1.3$\,keV$^{-1}$.\\    

From the halo flux in one annulus, one can now estimate the \emph{expected} count enhancement in another annulus.
Because of the CCD gaps, we estimate the {expected} count enhancement in a fan region, e.g., from $20\arcsec$ to $40\arcsec$ (see Fig.~\ref{PSRWide} where a fan region up to $60\arcsec$ is shown).
From the enhancement of 31\,counts in the BEB and 27\,counts in the SEB in the respective annuli (Table~\ref{Gap}), we estimate  expected enhancements of $27\pm 5$\,counts (BEB) and $13\pm 4$\,counts (SEB) for the fan region.
From our MARX simulation and the background rate measurement, we expect that $94 \pm 11$\,counts (BEB) and $10 \pm 3$\,counts (SEB) actually come from the background and, negligibly, from the PSF wings. Hence, one would expect a \emph{total} count number of $121\pm 12$\,counts (BEB) and $23\pm 5$\,counts (SEB).
In the respective fan region of the actual data, however, we find $98 \pm 10$\,counts (BEB) and $15 \pm 4$\,counts (SEB), i.e., the measured difference between halo prediction and data is $23 \pm 16$\,counts (BEB) and $8 \pm 6$\,counts (SEB).
Though we find notably fewer counts than one would expect in presence of a dust halo, the strong background contribution renders the significance of the difference between prediction and actual data to $<2\sigma$ in both energy bands. Hence, a dust halo still cannot be excluded.
We note that such profile estimates are obviously biased with respect to the assumed dust grain properties and dust distribution along the line of sight. A sharper surface brightness profile could be produced, for example, by a localised dust clump very close to the pulsar. 
However, there is no obvious indication for such a dust clump in the direction of $\psr$ in 3D ISM models (see, e.g., \citealt{Lallement2014}).\\

As shown by \citet{Predehl1995}, one can assume $\tau_{\rm sca}({\rm 1\,keV})=S N_{\rm H,22}$ with $N_{\rm H,22}=N_{\rm H}/(10^{22}$\,cm$^{-2}$) and $S$ a constant on the order of 1 (e.g., $S\simeq 0.5$; \citealt{Predehl1995}). 
Aiming to constrain $S$, we use the measured BEB pulsar counts, $1858 \pm 43$ (Section~\ref{results}), the determined BEB counts of the enhancement in the annulus $4.88\arcsec-13.10\arcsec$, $31.2\pm 9.4$ from Table~\ref{Gap}, and $N_{\rm H,22}=0.034 \pm 0.001$ from the {\sl Chandra}/XMM fit of the pulsar spectrum. We obtain $S=10.5\pm 3.2$.  This value appears to be substantially larger than one expects for a dust scattering halo. Due to the low count number in the enhancement and the resulting large $3\sigma$ error of $S$, however, we cannot entirely exclude that a dust scattering halo is the only or main source of the found flux enhancement around \psr.\\

The $3\sigma$ asymmetry of X-ray count distribution in the $1\farcs{5}-4\arcsec$ region around the pulsar could be either a statistical noise feature or indication of a very faint and very compact PWN. Since the slight count enhancement is located in the direction of the proper motion, one could suspect a bow-shock PWN.
From optical observations, \citet{Mignani2010} measured the proper motion of $\psr$ as $\mu =42 \pm 5$ mas yr$^{-1}$. This translates into a transverse velocity $v_{\perp}=(70 \pm 8)$\,km\,s$^{-1}$ at a distance of 350\,pc. As this velocity is considerably larger than the typical sound speed in the ISM, $c_S \sim 10-30$\,km\,s$^{-1}$, one would indeed expect that the pulsar should create a bow-shock PWN. 
Using the transverse velocity as a lower limit to the {total} velocity, a typical ambient density of the ISM, ${\rho}_{\rm ISM} \sim 2 \times 10^{-24}$\,g\,cm$^{-3}$, and applying the stand-off radius $R_s$ formula (e.g., \citealt{Chatterjee2002}), 
$R^2_s=\dot{E}\, {(4 \pi c \, {\rho}_{\rm ISM} \, v^2_{\rm total})^{-1}}$,
one can estimate $R_s \lesssim 2.7 \times 10^{16}$\,cm, corresponding to $\lesssim 5\arcsec$ at $d=350$\,pc for the expected stand-off radius of a bow-shock PWN around \psr. Here, smaller values of $R_s$ would be expected if $v_{\rm total} > v_{\perp}$, i.e., if a significant radial velocity component is present.
Overall, the found asymmetry - if real - is consistent with, or slightly smaller than the expected bow shock dimensions.
However, given the faintness of this asymmetry emission and the preponderance of soft counts,  
background fluctuations are equally likely.\\

One could speculate about a model which could accommodate the different radial extents by assuming that $\psr$ has a significant radial velocity, i.e., the pulsar is moving towards us or away from us. 
From simulations of bow-shock nebulae, it was found that the radial extent of the cylindrical PWN tail can be estimated as $R_{tail} \sim 4 R_s$ (e.g., \citealt{Bucciantini2005}). From these models, one expects the shocked pulsar wind to flow with nearly relativistic velocities. Therefore, considering beaming and projection effects for a pulsar moving along the line of sight, one would expect to see a PWN with an extent of $\sim 4 R_s$ if the pulsar moves away from us and an extent of $\sim R_s$ if the pulsar moves toward us.
If the pulsar moves away from us with $v_{\rm total} \gg v_{\perp}$, we would expect a nearly round PWN.
In the case of $\psr$, additional radial movement away from us could 
explain the observed features: the  $4\farcs{9}-13\arcsec$ flux enhancement and the small $1\farcs{5}-4\arcsec$ asymmetry.\\ 

The estimated unabsorbed BEB flux of the enhancement (see Section~\ref{resextended}), corresponds to a luminosity of about $L^{\rm ext,2}_{\rm BEB} \sim 1\times 10^{29} d^2_{350}$\,erg\,s$^{-1}$ or $L^{\rm ext,4}_{\rm BEB}\sim 2 \times 10^{29} d^2_{350}$\,erg\,s$^{-1}$  with $d_{350}=d/350$\,pc for assumed photon indices of 2 or 4, respectively.
If the extended emission is mainly due to a PWN, the corresponding efficiency $\eta_{\rm PWN}=L^{\rm PWN} / \dot{E}$ is between $3 - 8\times 10^{-6}$. This efficiency is lower than for typical PWN around young pulsars, but in fact very similar to the value of Geminga's PWN (e.g., Figure 5 by \citealt{Kargaltsev2008}).
The found prevalence of soft counts could be either due to an unusually soft PWN spectrum or just due to the faintness of the PWN (for a typical power law slope, more flux contribution is expected from soft than from hard photons). 
The current count numbers are too low to reasonably constrain the spectrum of the flux enhancement and disentangle putative PWN and scattering dust halo contributions.
As explained above, a major contribution from a dust scattering halo seems unlikely, but cannot entirely be excluded.

\subsection{Pulsar}
\label{psrdiscuss}
An obvious reason for the differences between the \emph{XMM-Newton} EPIC and \emph{Chandra} ACIS results for the pulsar spectrum could be  cross-calibration effects.
Cross-calibration between ACIS and EPIC has been investigated by \citet{Schellenberger2015,Kettula2013,Tsujimoto2011,Nevalainen2010}.
These studies compare, e.g., spectral fit results for absorbed ($N_H \sim$ a few $10^{20}$\,cm$^{-2}$) relatively hard sources, usually galaxy clusters. 
They consistently find higher temperatures for ACIS than for EPIC, in particular for EPIC-pn. In their soft X-ray band ($0.2-2.0$\,keV), \citet{Kettula2013} reported 18\% higher ACIS (APEC model) temperatures than obtained with EPIC-pn. We refer to Figures 5, 6 and 11 by \citet{Schellenberger2015} for an illustration of the higher-ACIS-temperature effect. In addition, Figure 13 by \citet{Schellenberger2015} shows that ACIS results usually imply a larger $N_H$ than pn results.
For soft X-ray sources, there are no comparable systematic studies to our knowledge.
An example for a comparison between soft spectra from  ACIS and EPIC (with sufficient number of counts) is the study of the isolated neutron star RBS 1223 by \citet{Haberl2003}. In that work, the ACIS-S\footnote{Note that there are slight differences for ACIS-S and ACIS-I comparisons (e.g., Fig. 2 by \citealt{Schellenberger2015})} data from 2000 and the EPIC data from 2001 and 2003 were fit with an absorbed blackbody with a broad gaussian absorption line. The BB temperature was only slightly higher for ACIS-S ($87.8 \pm 1.0$ versus $85.8\pm 0.5$ for EPIC), and the inferred $N_H$ limit was smaller ($<1.6 \times 10^{20}$\,cm$^{-2}$ versus $<4.1 \pm 0.1 \times 10^{20}$\,cm$^{-2}$ for EPIC-pn)  than the EPIC values. 
Hence, for these particular observations of a mildly absorbed soft source, one obtains consistent BB temperatures, and consistent or \emph{smaller} $N_H$ values for ACIS with respect to EPIC -- in contrast to the findings in the systematic studies of the galaxy clusters. 
Without a systematic study of mildly absorbed soft sources it is therefore difficult to accurately assess the actual cross-calibration effects for a  comparison of ACIS\,/\,EPIC spectral fit results.\\

For RBS1223, \citet{Haberl2003} found a difference of about 10\% in the absorbed flux (in the same direction as our results).  
\citet{Nevalainen2010} (their Figure 20, Table 11) reported that flux differences between ACIS and pn results were only 2\% in their soft band ($0.2-2.0$\,keV) using ACIS data obtained before 2008. 
For \psr, the flux difference between the EPIC and ACIS results is 29\%  and 47\% for tied ($N_H$, PL photon index) and free parameter fits, respectively (Table~\ref{specfit}). Even considering errors, this appears to be much larger than a reasonable cross-calibration effect expected from previous analyses. 
Our \emph{Chandra} observations helped to evaluate the effect of contaminating neighbor sources on the flux. Inclusion of Source A in the ACIS-only fit of the 2BB$+$PL model resulted in an observed flux of 
$8.1^{+2.1}_{-1.1}\times 10^{-13}$\,erg\,cm$^{-2}$\,s$^{-1}$ in comparison to 
$8.0^{+2.1}_{-1.1} \times 10^{-13}$\,erg\,cm$^{-2}$\,s$^{-1}$ if this source was not included. 
Thus, one would expect the \emph{XMM-Newton} flux to be higher by only $1$\% (with a formal error of 37\%) if Source A was not subtracted (it was for MOS1/2). 
The formal errors are large due to uncertain spectral fit constraints. However, the corresponding increase in the count numbers of $4\pm3$\% (Source A has 77 counts, the pulsar has 1912; see Section~\ref{respsr}) support the finding that less than $10$\% of the flux difference can be attributed to inclusion of Source A.   
The other possible contaminating sources would have an even smaller (factor of 4) effect on the \emph{XMM-Newton} results. 
Hence, contaminating neighbor sources and a formal cross-calibration error of 10\% cannot account for an overall flux difference of 29\% or more. \\

With respect to the cross-calibration studies mentioned above, one needs to consider possible calibration changes over time, too. While EPIC-pn was found to be very stable over time \citep{Sartore2012}, this is not the case for ACIS. 
A contaminant has been accumulating on the optical-blocking filters of the ACIS detectors with a different rate over time. 
While the effect on the inferred spectral parameters of an X-ray source should be minimized by the contamination model implemented in the ACIS data reduction, an unavoidable uncertainty in the contamination correction can lead to a different measured flux and can offset the derived spectral parameters. A underestimated contamination, for example, would lead to a larger-than-real $N_H$ in the spectral fit results.
Recently, we discussed a 4\%--6\% decrease of the absorbed flux (and apparent slight $N_H$ increase) for the Central Compact Object in the Cas A supernova remnant which was measured with ACIS-I between 2006 and 2012 \citep{Posselt2013}. We showed that, in principle, the uncertainty in the ACIS calibration in a small energy range could be responsible for this small effect.
In the case of \psr, it is intriguing that the temperatures and emission area sizes of the ACIS-only fit (free and very large $N_H$ which could be partly due to the ACIS contamination) are much closer to the EPIC values than the numbers obtained if $N_H$ is tied to the much smaller EPIC value.
Currently, we cannot explain the surprisingly large flux difference between the ACIS and EPIC results. 
We suspect a cross-calibration issue, but this can only be probed with further X-ray observations, preferably with the time-stable EPIC-pn.\\

Our \emph{Chandra} data are also useful to assess the effect of contaminating neighbor sources on the measured photon index. 
As described in Section~\ref{respsr}, exclusion of the neighbor source A results in a slightly softer PL. In particular, using more restricted source extraction regions for the \emph{XMM-Newton} data together with the ACIS data, we find a slightly larger photon index than 
\citet{deluca2005} ($1.9 \pm 0.2$ versus $1.7 \pm 0.1$)  
The PL component of the optical spectrum of $\psr$ has a photon index of $2.05\pm 0.34$ \citep{Mignani2010}.
Though formally both components have similar photon indices, the new X-ray photon index does not change the fact that the extension of the X-ray PL component still overshoots the optical fluxes by a factor $\sim 4$. This is, however, similar to most other pulsars detected in the optical (e.g., \citealt{Mignani2011}).
We note that at {\sl Fermi} $\gamma$-ray energies, the photon index is harder, $\Gamma_{\gamma}=1.1 \pm 0.1$ (\citealt{Abdo2013}; see also Table~\ref{pulsars}), indicating an additional spectral break between the X-ray and the $\gamma$-ray spectral ranges. 

\subsection{Comparison with the PWN properties of other pulsars}
Of the 13 nearby ($<2$\,kpc) middle-aged rotation-powered pulsars that have sensitive public X-ray observations with {\sl Chandra} or XMM-$Newton$  (Table~\ref{pulsars}), only 6 have a prominent (relatively bright, extended) X-ray PWN, while 7 others have none or very compact faint ones.
The question arises what is the defining physical factor for the presence of a prominent X-ray PWN ?
One needs to consider the intrinsic pulsar properties, exterior environment factors, as well as sight-line dependent geometric effects.
Our selection for a restricted $\dot{E}$ range ensures that the spin-down energy alone cannot be the defining factor.
PSR\,J1741$-$2054, for example, has a factor of 3 smaller $\dot{E}$ than the {\textit{Three Musketeers}}, but a prominent X-ray PWN. 
Since the 13 pulsars have very similar spin-down properties, their inferred magnetic dipole fields have similar strengths. Thus, one would expect similar magnetospheres, capable of producing similar amount of accelerated charged particles.
The cutoff energy in $\gamma$-rays, another indicator of the availability of sufficiently accelerated particles in the magnetosphere, does not show a correlation with (non-)detections of PWNe either (Table~\ref{pulsars}).\\ 

The efficiency of particle acceleration in the magnetosphere can depend on the angle between the rotational and magnetic axis, the obliquity $\alpha$ (e.g., \citealt{Philippov2014,Tchekhovskoy2013}). 
While axis alignment in pulsars has been suggested on Myr timescales \citep{Young2010}, the pulsars in Table~\ref{pulsars} are too young for this effect to be relevant. Different axis orientations can produce different pattern of detections in radio, X-rays and $\gamma$-rays (see, e.g., \citealt{Abdo2013} for a discussion about different radio-loud/radio-quiet fractions of $Fermi$ pulsars with different $\dot{E}$).
For the considered middle-aged pulsars, there is no obvious correlation between the PWN (non-)detections and loudness or quietness in the radio and $\gamma$-ray frequency ranges.
Given the dependence of radio  and $\gamma$-ray detections on the respective beam widths and the angles between the direction of sight and the rotation axis, $\zeta$, the disentanglement of geometric factors is non-trivial. \citet{Pierbattista2015} used four different magnetosphere models (PC -- Polar Cap, SG -- Slot Gap, OG -- Outer Gap, OPC -- One Pole Caustic) to fit radio  and $\gamma$-ray  pulse profiles separately and together. Generally, all the models perform poorly, in particular if both wavelength ranges are considered together.
With respect to $\psr$ (very faint/compact PWN), we restrict the discussion on the influence of geometric factors to the exemplary comparison with the Geminga pulsar (very prominent PWN).\\

For the Geminga pulsar, the best fitting model used by \citet{Pierbattista2015}, the SG model, resulted in $\alpha=42^\circ$, $\zeta=51^\circ$. Even this model, however, poorly reproduces Geminga's $\gamma$-ray interpulse.
Previously, \citet{Watters2009} also obtained constraints on the geometry of the Geminga pulsar, but did not show the respective fitted pulse profiles. Using the OG-model they derived $\alpha=10^\circ-25^\circ$, $\zeta=85^\circ$; using a two-pole caustic magnetosphere model they obtained two solutions: $\alpha=30^\circ-80^\circ$, $\zeta=90^\circ$ and $\alpha=90^\circ$, $\zeta=55^\circ-80^\circ$. \citet{Malov1998} argued that the radio-quietness of Geminga indicates an aligned rotator.
While the current constraints on Geminga's $\alpha$ from its $\gamma$-ray pulse profile remain uncertain, the models seem to agree that  $\zeta>50^\circ$. If one assumes that the direction of proper motion and rotation axis are crudely aligned, this constraint on $\zeta$ means that  Geminga's total velocity is close to its measured transverse velocity.
Geminga's PWN consists of two large lateral tails and an axial tail (e.g., \citealt{Pavlov2010}).
The emission properties of the lateral tails resemble more those of jets than of a bow shock front, a fact which -- together with the axial tail -- seems to be indicative of a large $\alpha$  (Posselt et al., {\it{in preparation}}).\\
For \psr, \citet{Lyne1988} and \citet{Weltevrede2009} derive very similar obliquity values, $\alpha=75^\circ$, $\zeta=67^\circ-69^\circ$, from radio polarization measurements. 
Assuming that rotation axis and sight line are aligned, \citet{Malov2011} also used radio polarization measurements to obtain  $\alpha=8^\circ$ for \psr. 
Among the models used by \citet{Pierbattista2015} to describe the radio and $\gamma$-ray light curves of \psr, the OG-model and the PC-model gave the best fit results with $\alpha=77^\circ$, $\zeta=87^\circ$ and $\alpha=10^\circ$, $\zeta=7^\circ$, respectively. The radio pulse shape is poorly reproduced in the fits for both models.  
Overall, assuming an alignment of rotation and proper motion axes, the current constraints on $\psr$ seem to support either an oblique rotator, moving nearly perpendicular to the line of sight, or a nearly aligned rotator which moves along the line of sight.\\

The size of a bow shock created by the PWN around a pulsar depends not only on the pulsar energetics, but also on the pulsar velocity and the density of the ISM (as we discussed for $\psr$ in Section~\ref{disext}). The transverse velocity of \psr, $70$\,km\,s$^{-1}$, is low in comparison to the mean value of other normal pulsars, $246 \pm 22$\,km\,s$^{-1}$ \citep{Hobbs2005}. PSR\,J0659+1414 which also does not have a prominent PWN, has a similar low transverse velocity, while Geminga's transverse velocity is close to the mean value reported by \citet{Hobbs2005} (see Table~\ref{pulsars}). However, PSR\,J0358+5413 has a transverse velocity of only $58$\,km\,s$^{-1}$, yet also powers a prominent PWN \citep{McGowan2006}. Clearly, different velocities alone cannot explain the diverse picture of PWN detections around middle-aged pulsars.

The influence of the ISM density can be crudely judged by comparing $N_H$ values. 
For instance, different $N_H$ for similarly distant pulsars in the same sky area can indicate different ISM densities close to the pulsars themselves. 
The $N_H$ values of the pulsars in Table~\ref{pulsars} are all very similar, but of course local inhomogeneities in the ISM density may exist.
It is particularly interesting to compare Geminga and PSR\,J0659+1414. These pulsars are very close to each other, and the local 3D map of the ISM does not reveal any substantial density variations between these two pulsar positions (at a spatial resolution of $\sim 10$\,pc; \citealt{Lallement2014}).
Though $\psr$ is in a different part of the sky, its $N_H$ value and the density in its local 3D map of the ISM resemble those of Geminga and PSR\,J0659+1414.
The ISM conditions of the {\it{Three Musketeers}} appear to be similar enough that they cannot explain the pulsar's different PWN manifestations.\\

Overall, the governing factor for the different PWN properties of the middle-aged pulsars appears to be the obliquity of the pulsar in combination with the  inclination to the observer. 
With respect to the {\it{Three Musketeers}}, we suggest that $\psr$ and PSR\,J0659+1414 are nearly aligned rotators whose main velocity component is along the line of sight. In contrast, Geminga appears to be an highly oblique rotator which moves nearly perpendicular to the line of sight.  
Further improvements in pulsar magnetosphere models, e.g., those used by \citet{Pierbattista2015}, and the corresponding detailed analyses of $\gamma$-ray, X-ray and radio pulse shapes are needed to confirm this hypothesis.

\section{Conclusions}
The vicinity of PSR\,B1055$-52$ does not show a prominent X-ray PWN. A $4\sigma$ flux enhancement has been detected in the SEB in an annulus $4\farcs{9}-20\arcsec$ around the pulsar. Taking also into account the X-ray count distribution at larger separations from the pulsar and properties of known dust scattering halos, a PWN interpretation seems more grounded than a dust scattering halo, though we cannot entirely exclude a contribution from  the latter due to the low count numbers.  
We detected a slight ($3\sigma$) asymmetry in the count distribution $1\farcs{5}-4\arcsec$ around the pulsar. This asymmetry is most pronounced in the BEB and located in the direction of the pulsar proper motion. While this emission is consistent with a bow shock PWN component due to the pulsar's transverse velocity, we cannot rule out background count fluctuations or MARX model uncertainties as other causes for the detected asymmetry.\\

Comparing the properties of nearby middle-aged pulsars, we identified differences in geometrical axis orientations as the most likely reason for the fact that some of these pulsars have very prominent X-ray PWNe while their -- otherwise very similar -- cousins have none or very faint/compact PWNe.
The faint extended emission around $\psr$ is consistent with a bow shock around an aligned rotator moving nearly along the line of sight.\\

For \psr, we found a $30$\,\% (or larger) flux decrease between the 2000 XMM-$Newton$ and the 2012 {\sl{Chandra}} observation which could formally be described by smaller (but hotter) thermal emission areas. We suspect a cross-calibration issue, but this needs to be proven with further X-ray observations, preferably with the time-stable XMM-$Newton$ EPIC-pn.

\acknowledgments

The scientific results reported in this article are based on observations made by the Chandra X-ray Observatory.
Support for this work was provided by the National Aeronautics and Space Administration through Chandra Award Number GO2-13090X  issued by the Chandra X-ray Observatory Center, which is operated by the Smithsonian Astrophysical Observatory for and on behalf of the National Aeronautics Space Administration under contract NAS8-03060.

\bibliographystyle{apj}
\bibliography{B1055bib}

\end{document}